\documentclass[12pt]{iopart}

\usepackage{graphicx}

\usepackage{epsfig}

\begin{document}

\title[Magnetic Component of Quark-Gluon Plasma]{Magnetic Component of Quark-Gluon Plasma}

\author{Jinfeng Liao$^1$ \& Edward Shuryak$^2$}

\address{Department of Physics, SUNY Stony Brook, NY 11794, USA}
\ead{$^1$jiliao@ic.sunysb.edu ,
$^2$shuryak@tonic.physics.sunysb.edu}
\begin{abstract}
We describe recent developments of the "magnetic scenario" of
sQGP. We show that at $T=(0.8-1.3)T_c$  there is a dense plasma of
monopoles, capable of supporting metastable flux tubes. Their
existence allows to quantitatively explained the non-trivial
$T$-dependence of the static $\bar Q Q$ potential energy
calculated on the lattice. By molecular dynamics simulation we
derived transport properties (shear viscosity and diffusion
constant) and showed that the best liquid is given by most
symmetric plasma, with 50\%-50\% of electric and magnetic charges.
The results are close to those of the ``perfect liquid'' observed
at RHIC.
\end{abstract}


\section{The ``magnetic scenario" for sQGP}

Dirac famously showed in 1931 that quantum mechanics requires
inverse relation between magnetic coupling constant for monopoles
and electric coupling. 't Hooft and Polyakov found monopole
solution in 1974, for non-Abelian gauge theories with adjoint
scalars.  The ``dual superconductor'' idea of confinement by 't
Hooft and Mandelstam led to significant interest in QCD monopoles,
especially on the lattice. Guided by Motonen-Olive
electric-magnetic duality,  Seiberg and Witten have solved exactly
the ${\cal N}=2$ SUSY gauge theory in 1994, identifying properties
and dynamical role of monopoles. Inspired by these, we have
recently suggested in \cite{Liao:2006ry} that quark-gluon plasma
in $T=(1-1.5)T_c$ region is a liquid dominated by magnetic
monopoles. In this talk we present some results: others are
contained in the final talk by Shuryak \cite{Shuryak:2008pz}.
Different arguments for ``magnetic liquid'' were provided in
\cite{Chernodub:2006gu}.

In the past few years, quark-gluon plasma in $T=(1-2)T_c$ region has
been found to be strongly coupled, thus sQGP. Among others, two
major evidences include: (i) successful hydro descriptions of RHIC
measurements on radial and elliptic flows suggest a ``perfect
liquid'' behavior (see e.g. \cite{Shuryak:2008pz}) with low
viscosity and small diffusion, which requires strong coupling
among constituents; (ii) lattice results for the static $\bar Q Q$
potential\cite{Static_potential} in this region showed very
nontrivial $T$-dependence of the interaction absolutely beyond the
perturbative expectation. Below we show how in the newly suggested
magnetic scenario the monopoles explain the static potential and
make the ``perfect liquid''.

\section{Monopoles Explain the Static $\bar Q Q$ Potential at $T\approx T_c$}

\begin{figure}
\centerline{
\includegraphics[width=0.37\textwidth]{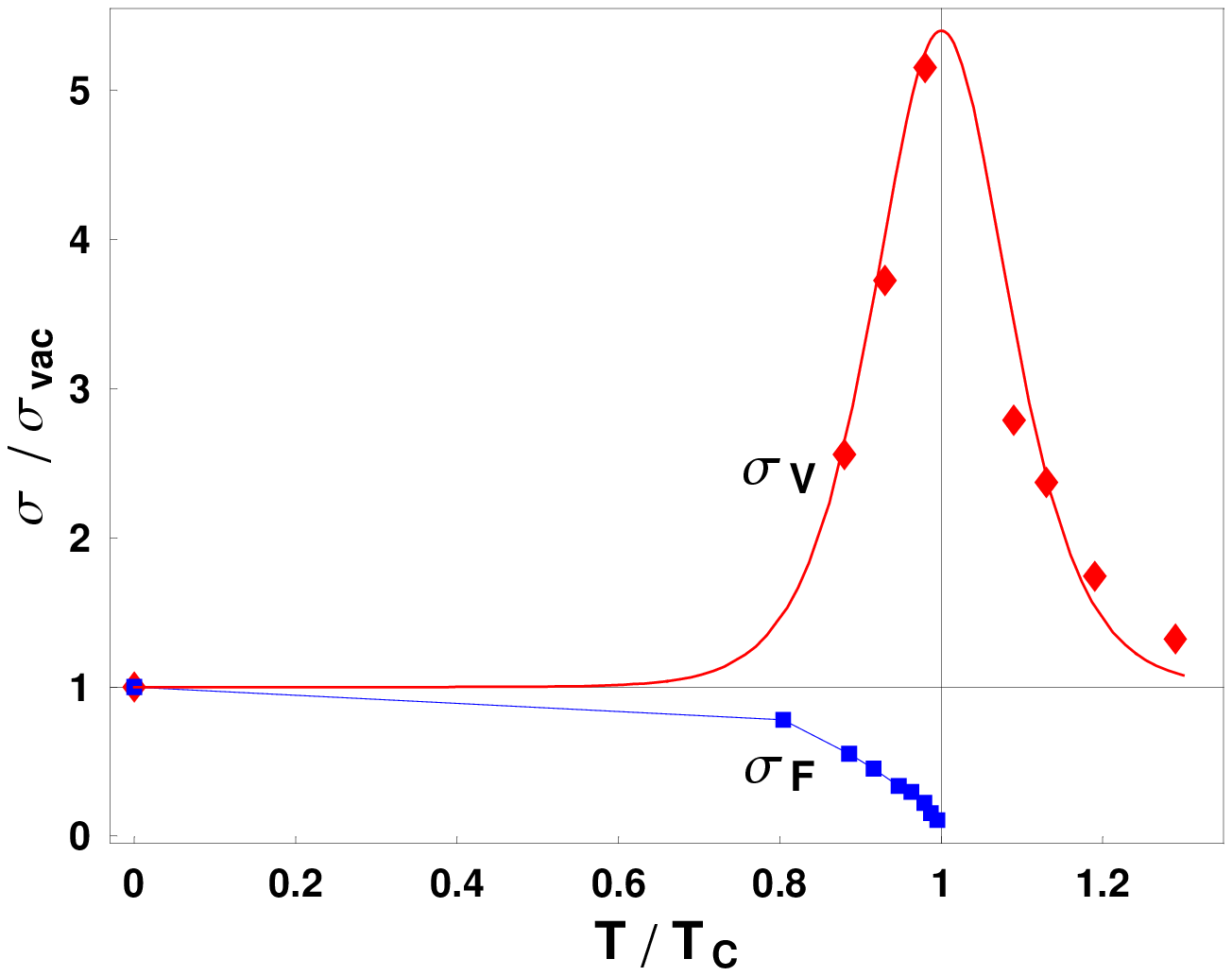}
  \hspace{0.1in}
  \includegraphics[width=0.32\textwidth]{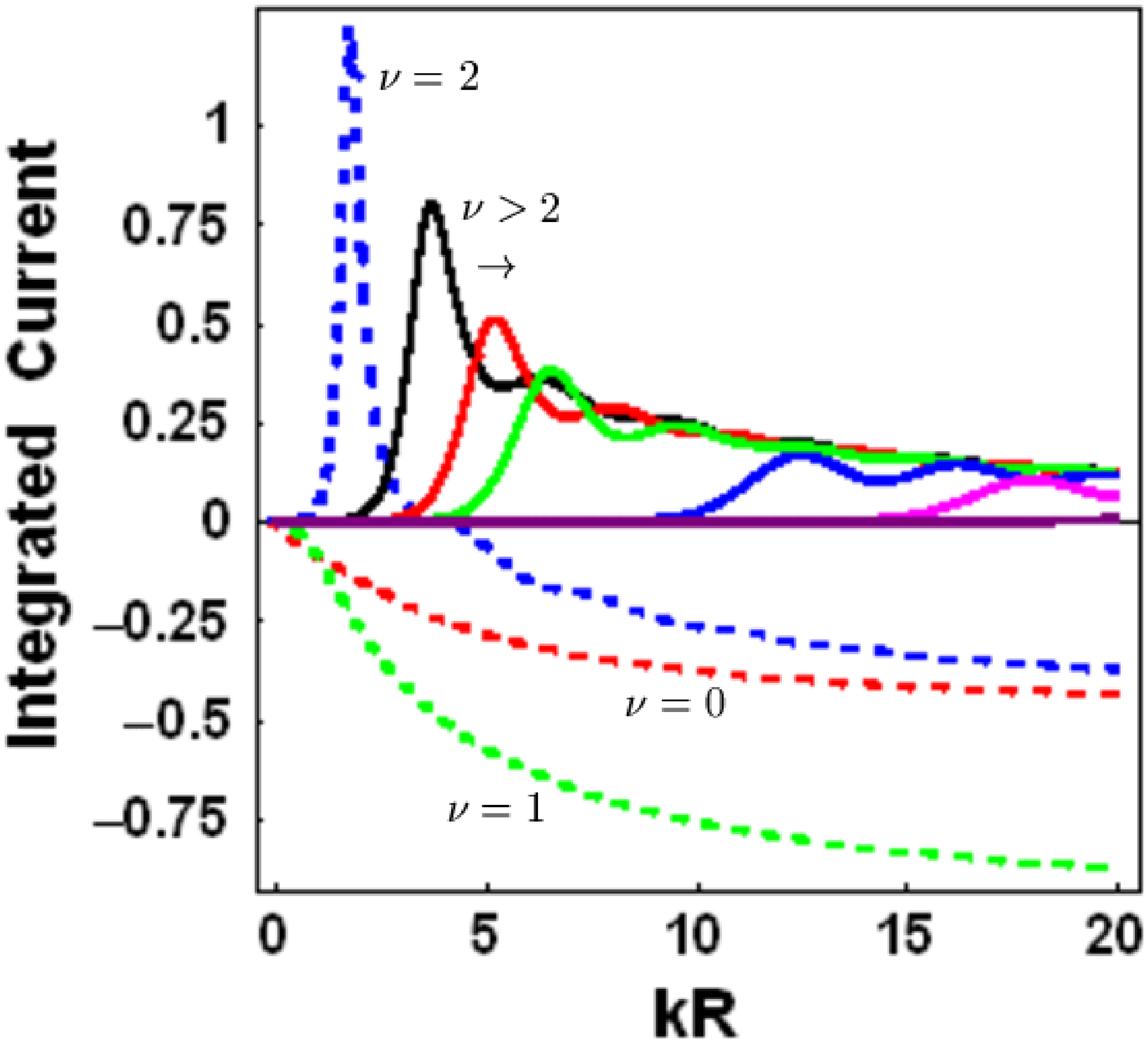}\\}
\centerline{\\
\includegraphics[height=0.28\textwidth]{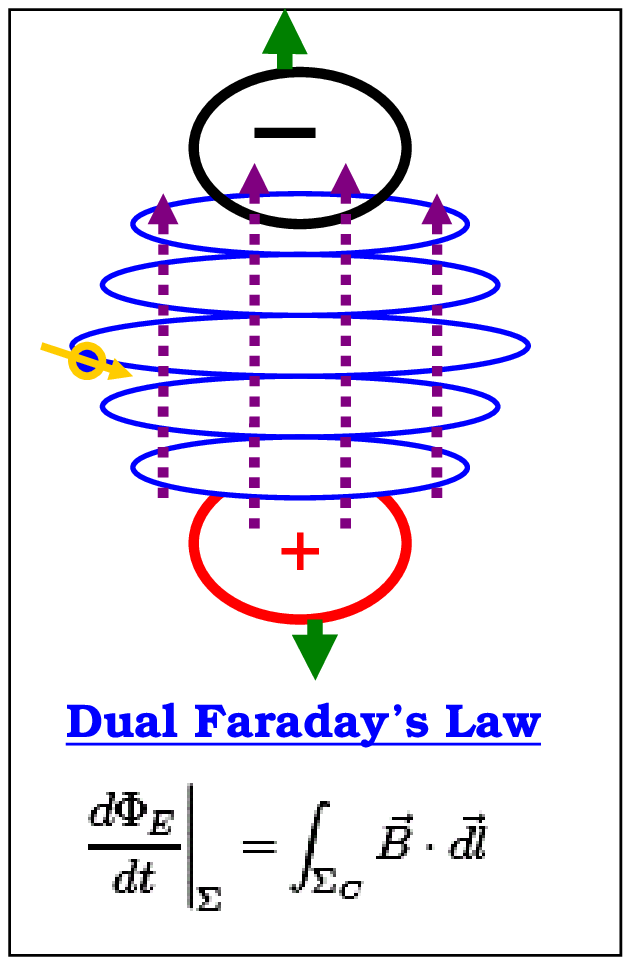}
 \hspace{0.25in}
 \includegraphics[width=0.40\textwidth]{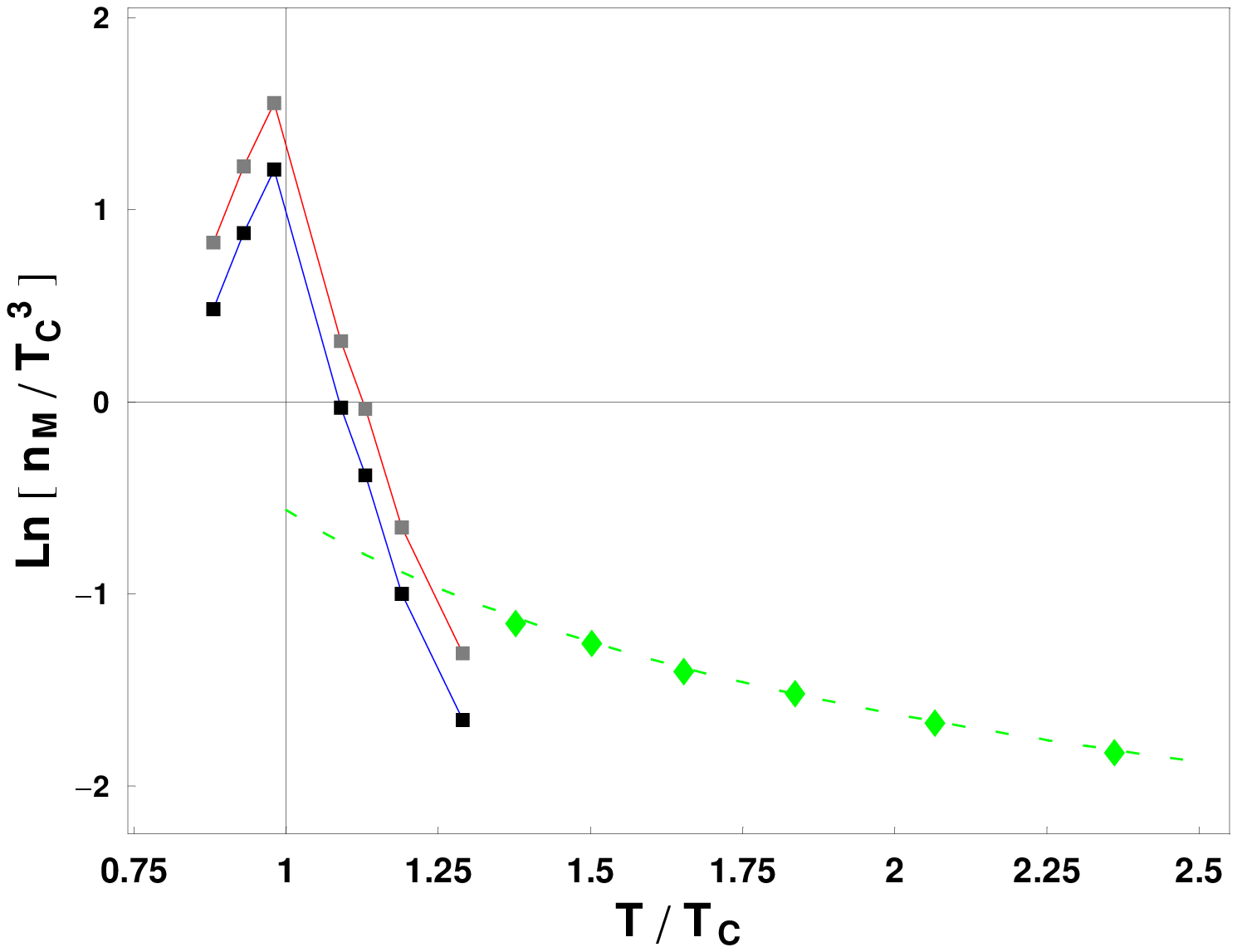}}
 \vspace{0.1in}
\caption{(a)(upper left) Effective string tension in free energy
$\sigma_F(T)$ and potential energy $\sigma_V(T)$ from lattice
results. (b)(upper right) Magnetic currents calculated from
monopole scattering on flux tube for various partial waves.
(c)(lower left) Schematic demonstration of magnetic solenoidal by
Dural Faraday's law. (d)(lower right) Monopole density $n_M$ in
$0.8-1.3T_c$ obtained from $\sigma_V$ via (\ref{density_tension}),
with diamond points and fitting curve in higher $T$ region from
\cite{D'Alessandro:2007su}.} \label{fig_static_potential}
\end{figure}

The static $\bar Q Q$ free energy $F(T,r)$ both below and above
$T_c$ has been calculated via lattice gauge
simulations\cite{Static_potential}. Furthermore the potential
energy $V(T,r)$ and entropy $TS(T,r)$ are also calculated via
$V=F-TdF/dT=F+TS$. Linear dependence on $r$, mostly in $0.5-1 fm$
region, is seen in free energy up to $T_c$. Remarkably in
potential energy $V(T,r)$, such linear dependence persists all the
way to $1.3T_c$, with $V(T,r\to \infty)$ (and $TS$) reaching
values as large as 4GeV at $T_c$! The $T$-dependence of the
effective string tensions $\sigma_F(T)$ and $\sigma_V(T)$, defined
as the slopes of linear parts in $F(T,r)$ and $V(T,r)$
respectively, are shown in Fig.\ref{fig_static_potential}(a).
While $\sigma_F$ vanishes at $T\rightarrow T_c$, $\sigma_V$ peaks
there, exceeding the vacuum string tension $\sigma_{vac}\approx
(426MeV)^2$ by about factor 5! Two questions have to be answered:
(i) why $\sigma_V$ survives to 1.3$T_c$ despite deconfinement;
(ii) what could be the origin of such large tension in the
potential energy and entropy around $T_c$ while $\sigma_F$ ceases
out. The electric objects are not relevant as they are confined
below $T_c$ and very heavy (thus rare) just above
$T_c$\cite{Liao:2005pa}, so the answers must lie in the magnetic
component.

The first question is answered in \cite{Liao:2007mj}:
 electric flux tube can be formed between the static $\bar Q
Q$ not only in a magnetic dual superconductor (as is the case
below $T_c$), but also in a {\it magnetic plasma} provided the
monopoles are {\it dense enough yet not too hot}. To see this, we
analytically calculated the quantum mechanic scattering of single
monopole with transverse momentum $k$ in the electric field $\vec
E$ of an infinitely long flux tube with size $R$ and also the
magnetic currents $\vec J_M$ from such scattering.
Fig.\ref{fig_static_potential}(b) shows the current contribution
from different angular momentum channel labelled by quantum number
$\nu$. The dual Maxwell equation relating $\vec E$ and $\vec J_M$
requires the current to be {\bf negative and strong enough} to
self-consistently hold the flux while
Fig.\ref{fig_static_potential}(b) shows only scattering in
$\nu=0,1$ channels are favorable and all higher partial waves are
``bad'' for flux tube. The plasma monopoles have typical $\nu \sim
k_T R$ thus to make flux tube we need the monopoles to be dense
and have $k_T$ as small as possible. We derived a quantitative
condition for mechanical stability of the flux tube :
\begin{equation} \label{critical_sQGP}
\frac{g^2}{4\pi}  (\frac{n}{T^3}) \ge 2.0  \left (
\frac{\bar{k}_T}{T} \right )^2   \frac{M}{T}
\end{equation}
The vanishing of $\sigma_V$ at $1.3T_c$ can be attributed to the
saturation of the above critical condition due to dropping
monopole density and rising $T$.

The second question is studied in \cite{LS_potential}. First of
all we identify slow/fast process (i.e. the process of separating
$\bar Q Q$ to a finite separation $L$) with free/potential energy
respectively. Supercurrent of condensed monopoles which has no
dissipation can not distinguish slow/fast process, while thermal
monopoles have finite relaxation time and do feel the difference.
At $T\approx T_c$ the originally dense condensate dies out as
signaled by vanishing $\sigma_F$, however these monopoles become
normal d.o.f and form a thermal ensemble with density $n_M$.
Suppose charges are separated very fast to $L$: a transverse loop
drawn in between will see significant time derivative of the
electric flux through it. Dual Faraday relation require the
circulation of magnetic field $\int \vec B \vec dl$ equal to this
change of flux, leading to a strong solenoidal magnetic current,
as illustrated in Fig.\ref{fig_static_potential}(c). Subsequently
the $normal$ currents by thermal monopoles damp out due to
collision with the bulk thermal matter, exchanging energy and
generating entropy till the equilibrium. In this way large amount
of heat $TS$ is generated which is associated with this pair of
charges, dissipating potential energy $V$ back to free energy
$F=V-TS$ which one finds for slow separation case. Based on this
picture we developed an analytic flux bag model and were able to
relate the effective tension $\sigma_V$ to the monopole density
\begin{equation} \label{density_tension}
\sqrt{\sigma_V(T)} = 3.88 \times \alpha_E^{1/6} \times
n_M(T)^{1/3}
\end{equation}
The $T$ dependence of the monopole density in $0.8-1.3T_c$ is
shown in Fig.\ref{fig_static_potential}(d): the results agree well
with the recent lattice results starting at
$1.3T_c$\cite{D'Alessandro:2007su}. We see monopole density
increases as $T \rightarrow T_c$ indicating they become light and
dominant in plasma and presumably reach condensation point at
$T_c$, below which the density drops quickly as thermal monopoles
turn into condensate. The density at $T>T_c$ suggests rapid
increase of magnetic screening toward $T_c$, which is also
consistent with lattice results \cite{Nakamura}.

\section{Monopoles Make the ``Perfect Liquid''}

\begin{figure}[t]
 \centerline{ \includegraphics[width=0.425\textwidth]{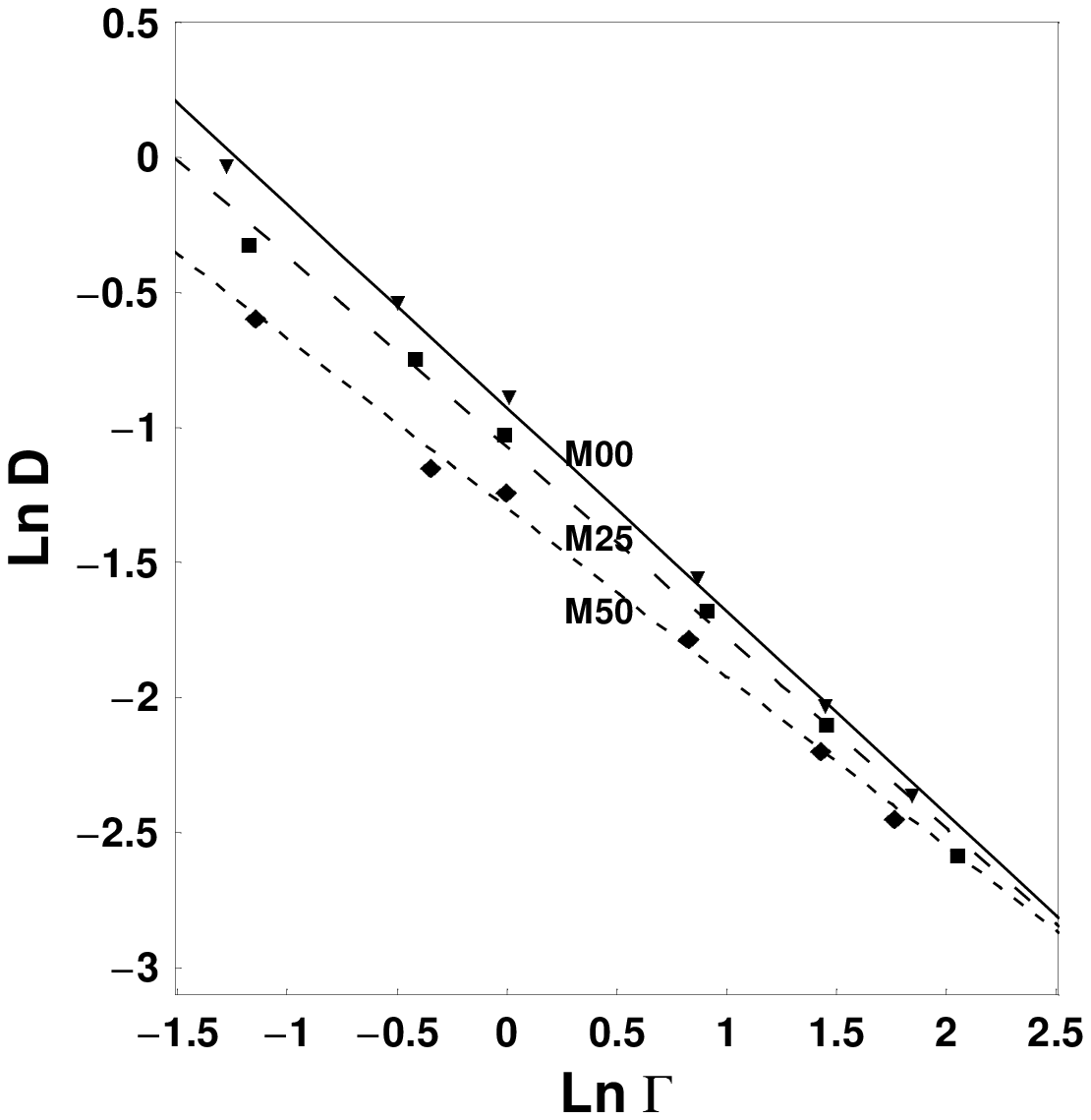}
  \hspace{0.1in}
  \includegraphics[width=0.44\textwidth]{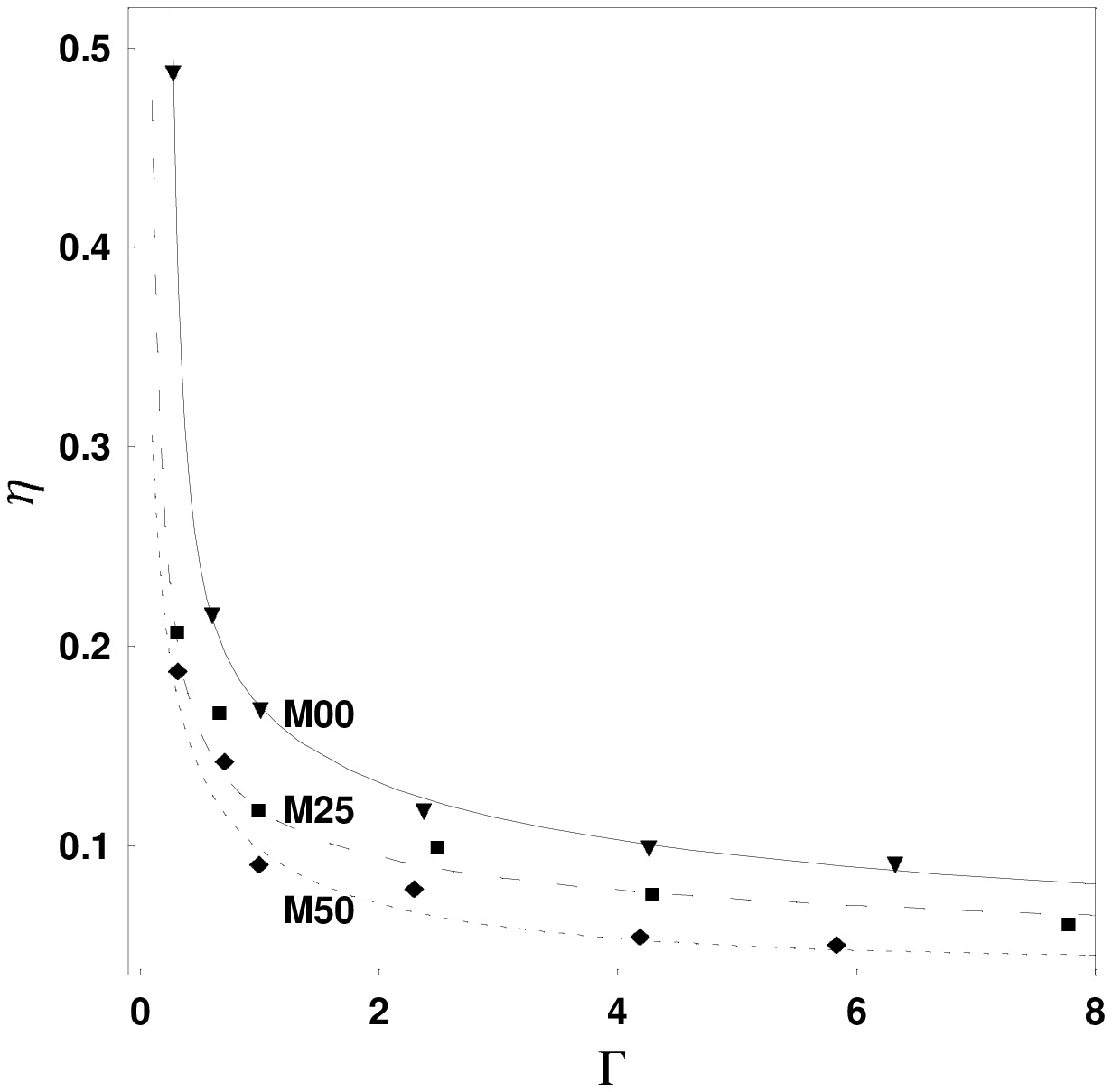}}
 \vspace{0in}
\caption{Diffusion constant $D$ (left) and viscosity $\eta$
(right) calculated at different plasma parameter $\Gamma$ for
M00(triangle), M25(square), and M50(diamond) plasma respectively
with the lines from fitting. } \label{fig_viscosity_diffusion}
\end{figure}

 It is important to check the consistency of the ``magnetic scenario'' for sQGP
 with the empirical discoveries at RHIC: i.e. whether
it can explain low viscosity and small diffusion constant.
 Molecular Dynamics (MD) simulations as powerful tools
for studying correlation functions and transport properties in
strongly coupled plasma have been used by us to explore (to our
knowledge, for the first time) a novel plasmas made of a mixture
of both electric and magnetic charges\cite{Liao:2006ry}. We used
standard Kubo formulae to calculate the shear viscosity $\eta$ and
diffusion constant $D$ in three settings: a pure electric plasma
(M00), a mixture with 25\% monopoles(M25), and a 50\%-50\%
mixture(M50). In Fig.\ref{fig_viscosity_diffusion} we show the
results $\eta$ and $D$ as functions of the key classical Coulomb
plasma parameter $\Gamma \equiv <potential>/<kinetic>\sim \alpha
\, n^{1/3}/T$ for all three settings. We found that at all
$\Gamma$'s, the most symmetric
 50\%-50\% mixture of charges and monopoles has
the smallest viscosity.
 The diffusion has a power law dependence on $\Gamma$,
for M50 we have $D=0.273/\Gamma^{0.626}$. With suitable mapping
from MD units to physical units of sQGP, the M50 plasma is found
to have viscosity and diffusion constant values very close to
those exacted from RHIC experiments: see transport summary Fig. in
 \cite{Shuryak:2008pz}. We conclude that the proposed plasma with both electric
and magnetic charges has the desired ``perfect liquid'' behavior.
More recently based on our MD as well as lattice results, a study
of monopole-anti-monopole equal-time correlator in
\cite{Liao:2008jg} further concluded that magnetic component of
quark-gluon plasma is also a liquid.

One may ask about the microscopic origin of such transport
properties in electric/magnetic plasma. We suggest an explanation
based on a ``magnetic bottle'' effect. (Invented by G.Budker in
1950's and routinely used in confined plasma fusion experiments.)
More specifically, in mixed plasma with  electric/magnetic
charges, we found that each charge can be trapped for long time
bouncing between surrounding {\it charges of the other kind}. The
Lorentz force leads to curling of the trajectory with decreasing
radius, forcing charges and monopoles collide more often than in
the gases made of only one type of particles, thus explaining the
``perfect liquid'' (to be  discussed
elsewhere\cite{Meta_trapping}).


\end{document}